\documentclass{elsart}
\usepackage{amsmath}
\usepackage{amsfonts}
\usepackage{amssymb}
\usepackage{graphics}
\usepackage{latexsym}
\usepackage{euscript}
\usepackage{supertabular}

\begin{document}
\numberwithin{equation}{section}
\numberwithin{table}{section}

\def\Bid{{\mathchoice {\rm {1\mskip-4.5mu l}} {\rm
{1\mskip-4.5mu l}} {\rm {1\mskip-3.8mu l}} {\rm {1\mskip-4.3mu l}}}}

\newcommand{\eL}{{\cal L}}
\newcommand{\J}{\textbf{J}}
\newcommand{\bP}{\textbf{P}}
\newcommand{\G}{\textbf{G}}
\newcommand{\K}{\textbf{K}}
\newcommand{\M}{{\cal M}}
\newcommand{\E}{{\cal E}}
\newcommand{\bu}{\textbf{u}}
\newcommand{\tr}{\mbox{tr}}
\newcommand{\norm}[1]{\left\Vert#1\right\Vert}
\newcommand{\abs}[1]{\left\vert#1\right\vert}
\newcommand{\set}[1]{\left\{#1\right\}}
\newcommand{\ket}[1]{\left\vert#1\right\rangle}
\newcommand{\bra}[1]{\left\langle#1\right\vert}
\newcommand{\ele}[3]{\left\langle#1\left\vert#2\right\vert#3\right\rangle}
\newcommand{\inn}[2]{\left\langle#1\vert#2\right \rangle}
\newcommand{\Real}{\Bid R}
\newcommand{\dmat}[2]{\ket{#1}\!\!\bra{#2}}

\begin{frontmatter}

\title{Generalized Euler Angle Parameterization for $U(N)$ with Applications to $SU(N)$ Coset Volume Measures.}

\author{Todd Tilma}
\address{Digital Materials Lab \\
Frontier Research System \\
The Institute of Physical and Chemical Research\\
Wako-shi, Saitama-ken, 351-0198 Japan}
\ead{tilma@riken.jp}

\author{E.C.G. Sudarshan}
\address{Center for Particle Physics \\
Physics Department \\
The University of Texas at Austin \\
Austin, Texas 78712-1081}
\ead{sudarshan@physics.utexas.edu}

\date{\today}

\begin{abstract}
In a previous paper \cite{Tilma2} an Euler angle 
parameterization for $SU(N)$ was given.  Here we present
a generalized Euler angle parameterization for $U(N)$.  
The formula for the calculation of the volume
for $U(N)$, $\mathbb{C}\mbox{P}^N$ as well as other $SU(N)$ and $U(N)$
cosets, normalized to this parameterization, will also be given.  
In addition, the mixed and pure state product measures
for $N$-dimensional density matrices under this parameterization will also be derived.  
\end{abstract}


\end{frontmatter}


\section{Introduction}

Having produced an Euler angle parameterization for $SU(N)$ we now turn
our attention to explicitly writing down the Euler parameterization
for the unitary group, $U(N)$ (which was hinted at in the $SU(N)$ work
of \cite{Tilma2,Tilma1}).
Recall that $U(N)$ is a subgroup of $GL(N,\mathbb{C})=GL_N(\mathbb{C})$, the group of
all complex $N \times N$ matrices with non-vanishing determinant and
requiring $2N^2$ parameters to represent.  In
this manner we can define $SU(N)$ to be a subgroup of $U(N)$, requiring $N^2-1$
parameters to represent, by adding the extra
condition that any element of $SU(N)$ has unit determinant.  We can
therefore expect that not only will the Euler parameterization of $U(N)$ be
easy to produce but
also the group volume, once we exploit
some simple group relationships between $SU(N)$ and $U(N)$.

The importance of such a parameterization and its corresponding volume
equation, beyond that discussed in \cite{Boya2}, is that it gives us the ability to calculate the measures
and volumes for general $N$-dimensional pure and mixed state density
matrices, as well as the volumes of the manifolds of operations on
pure and mixed states which produce entangled and 
separable states (which are directly related to the volume of 
separable and entangled states) without having to resort to extensive numerical
computations as in \cite{ZyczkowskiV1,ZyczkowskiV2}.


\section{Euler Parameterization of $U(N)$}

The idea behind the parameterization of $U(N)$ is straightforward.  Referring
to our previous work \cite{Tilma2}, for notations and details, as well as to 
Nakahara \cite{Nakahara}, Sattinger \cite{Sattinger}, 
and others we
know the following relationship holds between $SU(N)$, $U(N)$ and $\mathbb{C}\mbox{P}^N$:
\begin{equation}
\label{uandcpn}
\mathbb{C}\mbox{P}^N = \frac{{SU(N+1)}}{{U(N)}} = \frac{{SU(N+1)}}{{SU(N)} \times {U(1)}}.
\end{equation}
The $U(1)$ in the denominator of the second equality is the $U(1)$
element from the $SU(N+1)$ group in the numerator, which we know from
\cite{Tilma2,Tilma1}
to be:
\begin{equation}
\label{genu1}
{U(1)} \equiv {U(1)}_{{SU(N+1)}} = e^{i \lambda_{(N+1)^2-1} \beta}.
\end{equation}
Using the $SU(N)$ parameterization work done previously 
\cite{Tilma2}, 
we can write down the Euler
parameterization of $U(N)$ quite easily.

Recall from \cite{Tilma2,Tilma1,Sattinger,Herstein}  
we know we can write down a semi-direct sum for the Lie Algebra 
for $SU(N)$ as
\begin{equation}
L({SU(N)})=L({K}) \oplus L({P}),
\end{equation}
which yields a decomposition of the group,
\begin{equation}
{V}={K}\cdot {P},
\end{equation}
where $V \in SU(N)$.  From this work, we also know that $L({K})$ is comprised of the generators of the
$SU(N-1)$ subalgebra of $SU(N)$, and therefore
$K$ will be the $U(N-1)$ subgroup obtained by
exponentiating this subalgebra,
$\{ \lambda_1,\ldots,\lambda_{(N-1)^2-1} \}$, combined with
$\lambda_{N^2-1}$ and thus can be written as (see 
\cite{Tilma2,Tilma1,MByrd1} for examples)
\begin{equation}
\label{kdef}
{K(N-1)}=[{SU(N-1)}]\cdot e^{i\lambda_{N^2-1} \phi}
\end{equation}
where $[SU(N-1)]$ represents the $(N-1)^2-1$ term Euler angle 
parameterization of the $SU(N-1)$ subgroup.  

We are now ready to look at the $U(N)$ group in general.
For a $U \in U(N)$ we have from equations \eqref{uandcpn} and
\eqref{kdef} as well as from 
\cite{Tilma2}
\begin{equation}
\label{uNfinal}
{U} \equiv {K(N)} =[{SU(N)}]\cdot e^{i\lambda_{(N+1)^2-1} \beta},
\end{equation}
where
\begin{align}
[{SU(N)}]=&\;\biggl(\prod_{2 \leq k \leq N}A(k,j(N))\biggr) \cdot
\biggl(\prod_{2 \leq k \leq N-1}A(k,j((N-1))\biggr) \cdots
\biggl(A(2,j(2))\biggr)\nonumber \\
&\times
e^{i\lambda_{3} \alpha_{N^2-(N-1)}} \cdots e^{i\lambda_{(N-1)^2-1}
  \alpha_{N^2-2}} e^{i\lambda_{N^2-1} \alpha_{N^2-1}} \nonumber\\
=&\;\prod_{N \geq m \geq 2}\;\biggl(\prod_{2 \leq k \leq m}
A(k,j(m))\biggr) \nonumber \\ 
&\times e^{i\lambda_{3} \alpha_{N^2-(N-1)}} \cdots e^{i\lambda_{(N-1)^2-1}
  \alpha_{N^2-2}} e^{i\lambda_{N^2-1} \alpha_{N^2-1}},\nonumber \\
A(k,j(m))=&\; e^{i\lambda_{3}
  \alpha_{(2k-3)+j(m)}}e^{i\lambda_{(k-1)^2+1} \alpha_{2(k-1)+j(m)}},
  \nonumber \\
j(m) =&
\begin{cases}
0 \qquad &m=N,\\
\underset{0 \leq l \leq N-m-1}{\sum}2(m+l) \qquad &m \neq N. 
\end{cases}
\label{eq:suN}
\end{align}


\section{Volume of $U(N)$}

From 
\cite{Tilma2,Marinov2} 
the volume of $SU(N)$ is known to be
\begin{equation}
\label{eulermarinov}
V_{{SU(N)}}=2^{\frac{N-1}{2}}\pi^{\frac{(N-1)(N+2)}{2}}\sqrt{N}\prod^{N-1}_{k=1}
\biggl(\frac{1}{k!}\biggr).
\end{equation}
If we use equation \eqref{genu1}
then from 
\cite{Tilma2} 
we can define the following volume for ${U(1)}_{{SU(N+1)}}$.
\begin{eqnarray}
\label{Uvol}
V_{{U(1)}_{{SU(N+1)}}} \equiv V_{{U(1)}_{\lambda_{(N+1)^2-1}}}&=&(N+1)*\int_0^{\pi\sqrt{\frac{2}{(N+1)((N+1)-1)}}}d\alpha_{(N+1)^2+1}\nonumber \\
&=&\; \pi\sqrt{\frac{2(N+1)}{N}}. 
\end{eqnarray}
From equation \eqref{uNfinal} we can write
\begin{equation}
V_{{U(N)}}=V_{{SU(N)}}\times V_{{U(1)}_{\lambda_{(N+1)^2-1}}},
\end{equation}
and thus using equations \eqref{eulermarinov} and \eqref{Uvol} we have
\begin{align}
V_{{U(N)}}&=\;2^{\frac{N-1}{2}}\pi^{\frac{(N-1)(N+2)}{2}}\sqrt{N}\prod^{N-1}_{k=1}
\biggl(\frac{1}{k!}\biggr)*\pi\sqrt{\frac{2(N+1)}{N}} \nonumber \\
&=\;2^{\frac{N}{2}}\pi^{\frac{N(N+1)}{2}}\sqrt{N+1}\prod^{N-1}_{k=1}
\biggl(\frac{1}{k!}\biggr)
\label{eq:UNvol}
\end{align}
for $N \geq 2$.  Note that when $N=1$ we generate the volume for the 
$U(1)_{SU(2)}$ group element.  Since there can be many different
$U(1)$'s with different volumes, the fact
that $U(N)$, when $N=1$ gives the $SU(2)$ group element volume demands
that we limit the use of equation (\ref{eq:UNvol}) to $N\geq 2$.

With this information in hand, we can now look at the differential volume
elements, and corresponding volumes of the full range of $SU(N)$ and
$U(N)$ cosets that are of interest in physics, beginning with the
fundamental manifolds which define pure and mixed states.


\section{Differential Volume Elements for Pure and Mixed States}

Now that we have an Euler angle parameterization for both $SU(N)$
and $U(N)$, for $N\geq 2$, we are now in a position to look at the group
representations of pure and mixed states in terms of our parameterizations. 

In general, the manifold of pure states is given by the 
sequence of maps:
\begin{equation}
\label{gpstates}
{U(N-1)} \mapsto {SU(N)} \mapsto \mathbb{C}\mbox{P}^{N-1}.
\end{equation}
These are related to the ``Grassmannian'' manifolds, which are defined
as
\begin{equation}
\label{grasscpn}
\mathbb{C}\mbox{P}^{N-1}\equiv G(N,1) 
= \frac{{U(N)}}{{U(1)}\times {U(N-1)}} = \frac{{SU(N)}}{{U(N-1)}}.
\end{equation}
On the other hand, the manifold for mixed states (here for rank $N$ density matrices with non-degenerate and non-singular eigenvalues) 
is given by \cite{MByrd3Slater1,MByrd4All}
\begin{equation}
\label{msstates}
\EuScript{M}_{ms} = \Omega_{N-1} \times \frac{{SU(N)}}{({U(1)})^{N-1}}
\end{equation}
where $\Omega_{N-1}$ can be seen as the $(N-1)$-dimensional solid
angle (with appropriate ranges) derived from the eigenvalues of
a suitably parameterized $(N-1)$-dimensional sphere (see \cite{Boya2,MByrd4All}), 
and the factor $({U(1)})^{N-1}$ is the maximal torus
spanned by the exponentiation of the Cartan subalgebra of the group
\begin{align}
(U(1))^{N-1} &= U(1)_{SU(2)} \times U(1)_{SU(3)} \times \cdots \times
U(1)_{SU(N)} \nonumber \\
&= U(1)_{\lambda_3} \times U(1)_{\lambda_8} \times \cdots \times U(1)_{\lambda_{N^2-1}}.
\label{eq:msflagdef}
\end{align}
One may also notice that $\EuScript{M}_{ms}$ is stratified by noting that 
\begin{equation}
\label{stratified}
\frac{{SU(N)}}{({U(1)})^{N-1}} \cong \mathbb{C}\mbox{P}^{N-1}\ltimes 
\mathbb{C}\mbox{P}^{N-2} \ltimes 
\cdots  \ltimes \mathbb{C}\mbox{P}^{1},
\end{equation}
where the $\ltimes$, denotes the (possibly) non-trivial topological 
product of the spaces.  These cosets are called flag manifolds and the
given topological product  
follows from the fact that the $SU(N)$ groups are products of 
odd-dimensional spheres (see \cite{Boya2} and references within).

Now, in order to do any ``physically'' meaningful calculation on either
manifold we require their measures; measures that can be derived by
using the Euler angle parameterizations of $SU(N)$ and $U(N)$.  It is
to this question that we now turn our attention to.
\subsection{Pure State Measure}
We know that pure states are in $\mathbb{C}\mbox{P}^N$ and from the
previous sections that 
\begin{equation}
\mathbb{C}\mbox{P}^N = \frac{SU(N+1)}{{U(N)}} = \frac{{SU(N+1)}}{{SU(N)} \times {U(1)}_{{SU(N+1)}}}.
\end{equation}
Using the differential volume element for $SU(N)$  
from \cite{Tilma2} 
we can immediately write
down the pure state measure as
\begin{align}
dV_{ps} &= \frac{dV_{{SU(N+1)}}}{dV_{{SU(N)}} \times dV_{{U(1)}_{{SU(N+1)}}}} \nonumber \\
&= \frac{K_{{SU(N+1)}}d\alpha_{(N+1)^2-1}\ldots
d\alpha_{1}}{K_{{SU(N)}}d\alpha_{N^2-1}\ldots d\alpha_{1} \times d\alpha_{(N+1)^2-1}}\nonumber \\
&= \biggl(\prod_{2 \leq k \leq N+1}Ker(k,j(N+1))\biggr)
d\alpha_{2N} \ldots d\alpha_{1}.
\label{eq:dvpurestates}
\end{align}
where from 
\cite{Tilma2}
\begin{equation}
Ker(k,j(N+1))=
\begin{cases}
\sin(2\alpha_{2}) \quad &k=2, \\
\cos(\alpha_{2(k-1)})^{2k-3}\sin(\alpha_{2(k-1)}) \quad
&2<k<N+1, \\
\cos(\alpha_{2N})\sin(\alpha_{2N})^{2N-1} \quad
&k=N+1,
\end{cases}
\end{equation}
with the following ranges
\begin{gather}
0 \le \alpha_{1} \le \pi, \text{ and } 0 \le \alpha_{2} \le \frac{\pi}{2} \nonumber \\
0 \le \alpha_{2j} \le \frac{\pi}{2},\quad
0 \le \alpha_{2j-1} \le 2\pi \nonumber \\
\text{ for } 2 \le j \le N.
\label{eq:cpnranges1}
\end{gather}
Note that these ranges are from the \textit{covering} ranges for
$SU(N+1)$ and not from ${SU(N+1)}/{Z}_{N+1}$ which are used to calculate the
invariant volume for $SU(N+1)$ 
(see the appendices in \cite{Tilma2} for more details).
On the other hand, one \textit{could} use the
$SU(N+1)/Z_{N+1}$ ranges
\begin{gather}
0 \le \alpha_{2j} \le \frac{\pi}{2}, \quad
0 \le \alpha_{2j-1} \le \pi, \nonumber \\
\text{ for } 1 \le j \le N
\label{eq:cpnranges2}
\end{gather}
but then one would need to add a normalization factor
of $2^{N-1}$ in front of the product in equation
(\ref{eq:dvpurestates}) in order to generate the correct volume for
$\mathbb{C}\mbox{P}^N$.
\subsubsection{Example Calculation: Two Qubit Pure State Measure}

It is interesting to note that equation (\ref{eq:dvpurestates}) for
$N=3$ is \textit{equivalent} to the ``natural'' measure (referred to
in 
\cite{Tilma1}) 
derived from the
Hurwitz parameterization (see \cite{Zyczkowski1} and references within).  To begin, we define
a general vector of a random 4-dimensional unitary matrix $U(4)$ as
\begin{equation}
\label{hurwitzPSI}
\ket{\Psi(\boldsymbol{\eta})} = \begin{pmatrix}
\cos(\theta_3) \\
\sin(\theta_3)\cos(\theta_2)e^{i\phi_3} \\
\sin(\theta_3)\sin(\theta_2)\cos(\theta_1)e^{i\phi_2} \\
\sin(\theta_3)\sin(\theta_2)\sin(\theta_1)e^{i\phi_1}
\end{pmatrix}
\end{equation}
where $0 \leq \theta_i \leq \pi/2$ and $0 \leq \phi_i
\leq 2\pi$ ($i=1,2,3$), and $\boldsymbol{\eta} = \{\theta_i,\phi_i\}$.
From this vector one can calculate the corresponding Fubini-Study
metric (here given as in \cite{Resta})
\begin{equation}
g_{\mu \nu} = \frac{1}{2}(\EuScript{F}_{\mu \nu} +
\EuScript{F}_{\nu \mu}),
\end{equation}
where in this case
\begin{equation}
\EuScript{F}_{\mu \nu}(\boldsymbol{\eta}) =
\bra{\frac{\partial}{\partial \eta_\mu}\Psi(\boldsymbol{\eta})}(\Bid_4 -
\ket{\Psi(\boldsymbol{\eta})}\bra{\Psi(\boldsymbol{\eta})})
\ket{\frac{\partial}{\partial \eta_\nu}\Psi(\boldsymbol{\eta})},
\end{equation}
the square root of the determinant of which yields the invariant measure for
$\mathbb{C}\mbox{P}^3$ \textit{under this representation}:
\begin{eqnarray}
\label{dvpsfromhurwitz3}
dV_{ps} &=& \text{Det}[\sqrt{g}]\\
&=& \cos({{\theta }_1})\sin({{\theta }_1})\cos({{\theta }_2}){\sin({{\theta }_2})}^3
\cos({{\theta }_3}){\sin({{\theta }_3})}^5 d\theta_3 d\phi_3 \ldots
d\theta_1 d\phi_1.\nonumber
\end{eqnarray}
The \textit{equivalent} aspect of our statement comes in when one explicitly evaluates
equation (\ref{eq:dvpurestates}) for $N=3$:
\begin{align}
dV_{ps} &= \frac{dV_{{SU(4)}}}{dV_{{SU(3)}} \times dV_{{U(1)}_{{SU(4)}}}} \nonumber \\
&= \biggl(\prod_{2 \leq k \leq 4}Ker(k,j(4))\biggr)
d\alpha_{6} \ldots d\alpha_{1} \nonumber \\
&=
\sin(2\alpha_{2})\cos(\alpha_{4})^{3}\sin(\alpha_{4})\cos(\alpha_{6})\sin(\alpha_{6})^{5}
d\alpha_6 \ldots d\alpha_1
\nonumber \\
&=2\sin(\alpha_{2})\cos(\alpha_{2})\cos(\alpha_{4})^{3}\sin(\alpha_{4})\cos(\alpha_{6})\sin(\alpha_{6})^{5}
d\alpha_6 \ldots d\alpha_1
\label{eq:dvpurestatesforcp3}
\end{align}
where the ranges on the $\alpha_i$'s are from equation
(\ref{eq:cpnranges1}).  Obviously there's some contradiction between
this measure and the one given in equation \eqref{dvpsfromhurwitz3}
but any concern it may raise should be
eliminated in the following work.

To begin we note that equation (\ref{eq:dvpurestatesforcp3}) can also be derived in 
the following manner that follows the arguments found in \cite{MByrdp1}.
First we
define a pure state as
\begin{equation}
\label{altpurestate4}
\rho_d^\prime = \begin{pmatrix}
0 & 0 & 0 & 0 \\
0 & 0 & 0 & 0 \\
0 & 0 & 0 & 0 \\
0 & 0 & 0 & 1 
\end{pmatrix}
= \frac{1}{4}(\Bid_4 - \sqrt{6}\,\lambda_{15})
\end{equation}
and then apply a $U \in SU(4)$ to yield
\begin{equation}
\rho = U \rho_d^\prime U^\dagger = \frac{1}{4}(\Bid_4 -\sqrt{6}\, U\lambda_{15} U^\dagger).
\end{equation}
Recalling that a general two qubit density matrix has the form
\begin{equation}
\rho = \ket{\Phi(\boldsymbol{\alpha})}\bra{\Phi(\boldsymbol{\alpha})} = \frac{1}{4}(\Bid_4 + \sqrt{6}\,\mathbf{n}\cdot \boldsymbol{\lambda})
\end{equation}
we can therefore solve for the components of $\mathbf{n}$ and in turn $\Phi(\boldsymbol{\alpha})$, via 
evaluating $n_j = \Phi^\dagger \lambda_j \Phi$ for $j=1,\ldots,15$.  
Doing these calculations yields (dropping an overall multiplicative phase
term dependent on the $\lambda_{15}$ element of the Cartan subalgebra found in $U(3)$)
\begin{equation}
\label{eulerPHI}
\ket{\Phi(\boldsymbol{\alpha})} = \begin{pmatrix}
\sin(\alpha_6)\cos(\alpha_4)\cos(\alpha_2)e^{-i(\alpha_1 + \alpha_3 + \alpha_5)} \\
-\sin(\alpha_6)\cos(\alpha_4)\sin(\alpha_2)e^{i(\alpha_1 - \alpha_3 - \alpha_5)} \\
-\sin(\alpha_6)\sin(\alpha_4)e^{-i\alpha_5}\\
\cos(\alpha_6)
\end{pmatrix}.
\end{equation}
Calculating and taking the determinant of the Fubini-Study metric as before but now 
\textit{under this representation} yields
the following invariant measure for $\mathbb{C}\mbox{P}^3$
\begin{equation}
dV_{ps} = \sin(2\alpha_2)\sin(\alpha_4)\cos(\alpha_4)^3\sin(\alpha_6)^5\cos(\alpha_6).
\end{equation}

One can see that $\ket{\Phi(\boldsymbol{\alpha})}$ is similar to $\ket{\Psi(\boldsymbol{\eta})}$ but not equal. 
Therefore in comparing the two measures we can only note the
following:
\begin{enumerate}
\item{The factor of 2 in equation (\ref{eq:dvpurestatesforcp3}) equates to having the range of $\alpha_1$ run from
$0$ to $2\pi$ rather than its original range set $0 \leq \alpha_1 \leq
\pi$ given in equation (\ref{eq:cpnranges1}) thus allowing one to
\textit{conceptually} equate $\theta_i$ with $\alpha_{2i}$ and $\phi_i$ with
a functional form of the $\alpha_{2i-1}$s.}
\item{Equation \eqref{dvpsfromhurwitz3} can be generalized to 
$\mathbb{C}\mbox{P}^N$ (see \cite{Zyczkowski1})
\begin{equation}
\label{Zyckpsmeasure}
dV_{ps} = \prod_{k=1}^{N-1} \cos(\theta_k)\sin(\theta_k)^{2k-1}
d\theta_k d\phi_k
\end{equation}
which obviously \textit{does not} have the same form as equation (\ref{eq:dvpurestates}),
but due to the invariance of the integral
\begin{equation}
\int_0^{\frac{\pi}{2}} \sin(\xi)^m\cos(\xi) d\xi =
\int_0^{\frac{\pi}{2}} \sin(\xi)\cos(\xi)^m d\xi
\end{equation}
\textit{does} yield the same invariant volume (see the next section
and \cite{Zyczkowski1}).}
\end{enumerate}
Thus the difference between the two pure
state measures is just in the way one initially chooses the distribution of the
angles $\boldsymbol{\eta}$ and $\boldsymbol{\alpha}$ in the space
$\mathbb{C}\mbox{P}^3$ (and in $\mathbb{C}\mbox{P}^N$ in
general).  Since we are most concerned with unitary
operators in $SU(N)$ acting upon pure state density matrices and not within
the more general $U(N)$ group, we feel that our representation of the
pure state measure is more useful with regards to the overall Euler
parameterization of $SU(N)$ and $U(N)$ than the one given in equation
\eqref{Zyckpsmeasure}. 
\subsection{Mixed State Product Measure}
From equation \eqref{msstates} we can see that, in general, 
one can write down the mixed-state product measure for $\rho=U\rho_d U^\dagger$
as
\begin{equation}
\label{msmeasure}
dV_{ms} = d\mu \times d\biggl(\frac{G}{H}\biggr),
\end{equation}
where $d\mu$ defines a measure in the ($N-1$)-dimensional symplex of eigenvalues of $\rho_d$
and $d(G/H)$, where $G=SU(N)$ and $H={U(1)}_{{SU(2)}}\times {U(1)}_{{SU(3)}}\times \cdots
\times {U(1)}_{{SU(N)}}$, defines a ``truncated'' Haar measure which is responsible for the choice of eigenvectors of $\rho$ that ensures $d\mu$ is ``rotational invariant.''

Now as \cite{ZyczkowskiV1,ZyczkowskiV2,Zyczkowski1,SlaterC,SlaterB,SlaterA,Zyczkowski2}
and others have noted, $d\mu$ is defined via the probability distribution induced on the ($N-1$)-dimensional
symplex but there can be more than one possible $d\mu$ that is applicable for a given system since there
can be more than one usable probability distribution.
As Hall noted:
\begin{quote}
...[I]f [mixed states] described by density operators are allowed, the requirement
of unitary invariance (thus there is no preferred measurement basis for extracting information)
only implies that the probability measure over the set of possible states is a function of the density
operator eigenvalue spectrum alone.  Hence a unique probability measure can be specified only
via some further principle or restriction, to be motivated on physical or conceptual grounds \cite{Hall}.
\end{quote}
Therefore
\begin{quote}
An ensemble of general states of a quantum system is in general described by a probability measure
over the density operators of the system.  given that probability measures transform in the same way as
volume elements under coordinate transformations, and that volume elements are in general properties of metric
spaces, this suggests that the distribution of density operators corresponding to a ``minimal knowledge'' (i.\ e.\
most random ensemble of possible states) ensemble may be obtained from the normalized volume element induced by some
natural metric on the set of density matrices \cite{Hall}.
\end{quote}
Thus, the volume measure is defined by the choice of metric, and since the metric is invariant under unitary 
transformations, defining $d\mu$ comes down to determining which metric is the most appropriate in defining
a statistical distance between two density matrices; especially when one adds the additional requirement that the 
metric satisfy certain criteria for entanglement measures (see for example 
\cite{HorodeckiM,Donald,Stockton,Vedral/Plenio/Rippin/Knight,Vedral2} and references within).  

Since there are multiple choices for distance measures between two density matrices, and therefore $d\mu$, and since
we want to keep our discussion as general as possible, we shall defer further discussion on $d\mu$ to other papers
(for example those previously cited) and just use the most general form of $d\mu$ in the spirit equation 
\eqref{msstates} and given in \cite{ZyczkowskiV2,SlaterB,SlaterA} and
references within; the Dirichlet distribution:
\begin{equation}
d\mu = \frac{\Gamma(s_1 + \cdots + s_N)}{\Gamma(s_1)\cdots \Gamma(s_N)} \Lambda_1^{s_1 -1} \cdots 
\Lambda_{N-1}^{s_{N-1} -1}(1-\sum_{j=1}^{N-1}\Lambda_j)^{s_N -1} d\Lambda_1 \ldots d\Lambda_{N-1}
\end{equation}
where $\sum_{}^{} \Lambda_j =1$ and $1 > \Lambda_j > 0$ are just the eigenvalues of $\rho_d$.  The Dirichlet distribution provides a means of expressing quantities that vary randomly, independent of each other, yet obeying the condition that there sum remains fixed.  In our case, $s_j \equiv s > 0$ thus
\begin{align}
d\mu &= \frac{\Gamma(Ns)}{N\Gamma(s)}\,\Lambda_1^{s-1} \cdots \Lambda_{N}^{s-1} d\Lambda_1 \ldots d\Lambda_N \nonumber \\
&= \alpha_s \Lambda_1^{s-1} \cdots \Lambda_{N}^{s-1} d\Lambda_1 \ldots d\Lambda_N,
\label{eq:dmu}
\end{align}
where the ranges for the $\Lambda_j$, we conjecture, are equal to
\begin{equation}
\label{dmuranges}
1 \geq \Lambda_N \geq \frac{1}{N} \text{ and } 0 \leq \Lambda_2, \ldots, \Lambda_{N-1} \leq \frac{1}{N}.
\end{equation} 
These ranges disagree with those given by Slater in 
\cite{SlaterNew} for $N=4$ 
but under integration, the difference between those in \cite{SlaterNew} and ours is a multiplicative
factor of $N$ upon the kernel.  The benefit of the ranges given here is that they are easily generalized, while those in 
\cite{SlaterNew} are not.

With $d\mu$ so defined, we are now free to look at the 
flag manifold $G/H$. 
By using equations \eqref{genu1} and (\ref{eq:suN}) we can see that the coset $G/H$ can be expressed as
\begin{align}
\frac{G}{H} &= \frac{{SU(N)}}{{U(1)}_{{SU(2)}} \times {U(1)}_{{SU(3)}} \times
  \ldots \times {U(1)}_{{SU(N)}}} \nonumber \\
&= \frac{\prod_{N \geq m \geq 2}\;\biggl(\prod_{2 \leq k \leq m}
A(k,j(m))\biggr)e^{i\lambda_{3} \alpha_{N^2-(N-1)}} \cdots
  e^{i\lambda_{(N-1)^2-1}\alpha_{N^2-2}} e^{i\lambda_{N^2-1}
  \alpha_{N^2-1}}}{e^{i\lambda_{3} \alpha_{N^2-(N-1)}} \cdots e^{i\lambda_{(N-1)^2-1}\alpha_{N^2-2}}e^{i\lambda_{N^2-1}\alpha_{N^2-1}}}\nonumber \\
&= \prod_{N \geq m \geq 2}\;\biggl(\prod_{2 \leq k \leq m}
A(k,j(m))\biggr),
\end{align}
where $A(k,j(m))$ is defined in equation (\ref{eq:suN}).  This
coset representation comes from the following observation; it allows
us to write down the ``truncated'' Haar measure $d({G}/{H})$ as
\begin{align}
d\biggl(\frac{G}{H}\biggr)&= d\biggl(\frac{{SU(N)}}{{U(1)}_{{SU(2)}}\times {U(1)}_{{SU(3)}} \times \cdots \times {U(1)}_{{SU(N)}}}\biggr) \nonumber \\
&= \frac{dV_{{SU(N)}}}{dV_{{U(1)}_{{SU(2)}}} \times dV_{{U(1)}_{{SU(3)}}} \times
  \ldots \times dV_{{U(1)}_{{SU(N)}}}} \nonumber \\
&= \frac{K_{{SU(N)}}d\alpha_{N^2-1}\ldots
  d\alpha_{1}}{d\alpha_{N^2-(N-1)}\ldots d\alpha_{N^2-1}}
\nonumber \\
&= K_{{SU(N)}}d\alpha_{N(N-1)}\ldots d\alpha_{1}
\label{eq:dghmeasure}
\end{align}
where from 
\cite{Tilma2}
\begin{align}
K_{{SU(N)}}&=\prod_{N \geq m \geq 2}\;\biggl(\prod_{2 \leq k \leq
  m}Ker(k,j(m))\biggr), \nonumber \\
Ker(k,j(m))&=
\begin{cases}
\sin(2\alpha_{2+j(m)}) \quad &k=2, \\
\cos(\alpha_{2(k-1)+j(m)})^{2k-3}\sin(\alpha_{2(k-1)+j(m)}) \quad
&2<k<m, \\
\cos(\alpha_{2(m-1)+j(m)})\sin(\alpha_{2(m-1)+j(m)})^{2m-3} \quad
&k=m,
\end{cases}
\end{align}
and $j(m)$ is from equation (\ref{eq:suN}).  

Now the ranges for the
$\alpha$'s can again be either the \textit{covering} ranges defined for the
first $N(N-1)$ $\alpha$'s of the Euler parameterization of $SU(N)$
(see \cite{Tilma2})
\textit{or} the $SU(N)/Z_N$ ranges
\begin{gather}
0 \le \alpha_{2j} \le \frac{\pi}{2}, \quad
0 \le \alpha_{2j-1} \le \pi, \nonumber \\
\text{ for } 1 \le j \le \frac{N(N-1)}{2},
\label{eq:psranges1}
\end{gather}
which would necessitate adding a normalization factor of
$2^{(N-1)(N-2)/2}$ to $K_{{SU(N)}}$ in equation (\ref{eq:dghmeasure}).

Depending on which set of ranges are used, a general mixed-state product measure can thus be written as 
\begin{eqnarray}
\label{dvmixedstates}
dV_{ms} &=& \alpha_s \Lambda_1^{s_1 -1} \cdots \Lambda_{N-1}^{s_{N-1} -1}(1-\sum_{j=1}^{N-1}\Lambda_j)^{s_N -1} 
d\Lambda_1 \ldots d\Lambda_{N-1} \nonumber \\
&&\times
\xi \cdot K_{{SU(N)}} d\alpha_{1}\ldots d\alpha_{N(N-1)},
\end{eqnarray}
where the $\Lambda_i$ are the \textit{non-zero} eigenvalues of the corresponding
$N$-dimensional diagonal density matrix $\rho_d$ (see 
\cite{Tilma2,ZyczkowskiV1,ZyczkowskiV2,MByrd3Slater1,Zyczkowski1,SlaterNew,Slater2} for
more details) and $\xi$ is the necessary normalization constant (equal
to 1 if one uses the covering ranges for $SU(N)$ and equal to
$2^{(N-1)(N-2)/2}$ if one uses the generic $SU(N)/Z_{N}$ coset ranges used in calculating
the group volume 
\cite{Tilma2,Tilma1}.
\subsubsection{Example Calculation: Two Qubit Mixed State Product Measure}

For two qubits, equations \eqref{msmeasure}, (\ref{eq:dmu}), and (\ref{eq:dghmeasure}) yield
\begin{align}
dV_{ms} =&\, d\mu \times d\biggl(\frac{SU(4)}{U(1)_{SU(2)} \times U(1)_{SU(3)} \times U(1)_{SU(4)}}\biggr) \nonumber \\
=&\, \alpha_s \Lambda_1^{s-1}\Lambda_2^{s-1}\Lambda_3^{s-1}(1-\sum_{i=1}^3 \Lambda_i)^{s-1}d\Lambda_1 \ldots d\Lambda_3 \times \xi \cdot K_{SU(4)}d\alpha_{12} \ldots d\alpha_1 \nonumber \\
=&\, \alpha_s \Lambda_1^{s-1}\Lambda_2^{s-1}\Lambda_3^{s-1}\Lambda_4^{s-1}d\Lambda_1 \ldots d\Lambda_4 \nonumber \\
&\times \xi \cdot \sin(2\alpha_{2})\sin(\alpha_{4})\cos(\alpha_{4})^3\sin(\alpha_{6})^5\cos(\alpha_{6}) \nonumber \\
&\times \sin(2\alpha_{8})\sin(\alpha_{10})^3\cos(\alpha_{10})\sin(2\alpha_{12})d\alpha_{12}\ldots d\alpha_1
\end{align}
where we have used 
the $SU(4)$ differential volume element from \cite{Tilma1}
in the last step.  The ranges of integration for the $\alpha_i$ parameters has already been discussed; ideally they should be the \textit{covering} ranges for $SU(4)$ 
from \cite{Tilma1}
so that $\xi=1$.  As for the
ranges on $d\mu$, recall that for two qubits, $\rho_d$ is given by 
\cite{Tilma2,Tilma1}
\begin{equation}
\rho_d = \left( \begin{smallmatrix}
\sin^2(\theta_1)\sin^2(\theta_2)\sin^2(\theta_3) & 0 & 0 & 0\\
0 & \cos^2(\theta_1)\sin^2(\theta_2)\sin^2(\theta_3) & 0 & 0\\
0 & 0 & \cos^2(\theta_2)\sin^2(\theta_3) & 0 \\
0 & 0 & 0 & \cos^2(\theta_3) 
\end{smallmatrix} \right)
\end{equation}
where
\begin{gather}
\begin{alignat}{3}
\frac{\pi}{4} \le \theta_1 \le \frac{\pi}{2}, \quad &
\cos^{-1}(\frac{1}{\sqrt{3}}) \le \theta_2 \le \frac{\pi}{2}, \quad &
\frac{\pi}{3} \le \theta_3 \le \frac{\pi}{2},
\end{alignat}
\end{gather}
thus we have
\begin{equation}
1 \geq \Lambda_4 \geq \frac{1}{4} \quad 0 \leq \Lambda_1 \leq \frac{1}{4}
\quad 0 \leq \Lambda_2 \leq \frac{1}{4} \quad 0 \leq \Lambda_3 \leq \frac{1}{4}
\end{equation}
for the ranges of integration on $d\mu$.  Notice that one could have
also used equation \eqref{dmuranges} with $N=4$ to achieve these
ranges, but it is instructive to see their explicit derivation.

\section{Volume of $\mathbb{C}\mbox{P}^N$ and $SU(N)/(U(1))^{N-1}$}

Now we are in a position to give two different methods for calculating
the volume for the pure and mixed state manifolds of general
$N$-dimensional quantum systems.  The pure state manifold
volume is quite simple and already well known; it is the volume of
$\mathbb{C}\mbox{P}^N$, 
while the mixed state manifold volume, as we have seen in the previous
section, is the
product of two different measures - one of which is dependent on the
initial distribution of states on the ($N-1$)-dimensional symplex.
Therefore in the mixed state case we shall only worry about
calculating the volume contribution
from the second measure; the ``truncated'' Haar measure, since the volume from the
$N-1$ symplex measure can be equated to a general multiplicative
constant determined by the initial distribution of states on the $N-1$ symplex (see for example \cite{SlaterNew}):
\begin{equation}
V_{\text{Mixed States}} = V_{\text{Symplex}} \times
  V_{SU(N)/(U(1))^{N-1}} = \omega V_{SU(N)/(U(1))^{N-1}}.
\end{equation}
For example, for the two qubit case described previously, 
a naive calculation of $\omega$ can be seen to be equal to:
\begin{align}
\omega &= \alpha_s \int_{\frac{1}{4}}^1 \int_0^{\frac{1}{4}} \int_0^{\frac{1}{4}} \int_0^{\frac{1}{4}} \Lambda_1^{s-1} \Lambda_2^{s-1}
\Lambda_3^{s-1} \Lambda_4^{s-1} d\Lambda_1 d\Lambda_2 d\Lambda_3 d\Lambda_4 \nonumber \\
&= \frac{\Gamma(4s)}{4\Gamma(2)}\biggr(\frac{4^{-4s}(-1+4^s)}{s^4}\biggl)
\label{eq:2qbomega}
\end{align}
for when $s > 0$ (note that the integration ranges on $\Lambda_4$ were reversed in order to keep $\omega$ positive).
\subsection{Volume of $\mathbb{C}\mbox{P}^N$}
Using the results for $U(N)$ we can immediately write down the general volume for $\mathbb{C}\mbox{P}^N$.  Using
equations \eqref{uandcpn}, \eqref{eulermarinov} and (\ref{eq:UNvol}) we have
\begin{align}
V_{\mathbb{C}\text{P}^N}&=\frac{V_{{SU(N+1)}}}{V_{{U(N)}}}
=\frac{2^{\frac{N}{2}}\pi^{\frac{N(N+3)}{2}}\sqrt{N+1}\prod^{N}_{k=1}
\biggl(\frac{1}{k!}\biggr)}{2^{\frac{N}{2}}\pi^{\frac{N(N+1)}{2}}\sqrt{N+1}\prod^{N-1}_{k=1}
\biggl(\frac{1}{k!}\biggr)}\nonumber \\
&=\frac{\pi^N}{N!}.
\label{eq:CPNvol}
\end{align}
We see that this result also comes from the integration of equation (\ref{eq:dvpurestates})
over the ranges given in equation (\ref{eq:cpnranges1}) or equation (\ref{eq:cpnranges2})
\begin{equation}
\idotsint\limits_{\alpha\; ranges} \biggr( \prod_{2\le k \le N+1} Ker(k, j(N+1)) \biggl) d\alpha_{2N} \ldots d\alpha_1 = 2^{N-1}\pi^N \prod_{2\le k \le N+1} \mathbb{V}(k,N+1) 
\end{equation}
where $\mathbb{V}(k,N+1)$ is, from 
\cite{Tilma2},
\begin{gather}
\mathbb{V}(k,N+1)=
\begin{cases}
1 \quad &k=2, \\
\frac{1}{2(k-1)} \quad &2<k\le N+1. 
\end{cases}
\end{gather}
Expansion of this product yields
\begin{equation}
\prod_{2\le k \le N+1} \mathbb{V}(k,N+1) = 1 \times \frac{1}{4} \times \frac{1}{6} \times \cdots \times \frac{1}{2N} = \frac{1}{2^{N-1}N!}
\end{equation}
which when multiplied by the $2^{N-1}\pi^N$ factor gives the volume
for $\mathbb{C}\mbox{P}^N$ as previously calculated.  One can 
also see that integration of equation \eqref{Zyckpsmeasure} using the
ranges given for $\boldsymbol{\eta}$ in equation \eqref{hurwitzPSI} 
will also generate equation
(\ref{eq:CPNvol}) (see \cite{Zyczkowski1} for example).

We should also note the remarkable results: 
\begin{align}
\sum_{n=0}^{\infty} \text{Vol}(\mathbb{C}\mbox{P}^n) =
\sum_{n=0}^{\infty} \frac{\pi^n}{n!} &= e^{\pi} \approx 23.147
\nonumber \\
\underset{k \rightarrow \infty}{\text{lim}}\,\prod_{n=0}^{k}
\text{Vol}(\mathbb{C}\mbox{P}^n) \rightarrow 0.
\label{eq:cpnobs}
\end{align}
Thus, in terms of our pure state manifold discussion, we can conclude
that as one increases the dimensionality of the system, there will
always be a non-zero pure state volume.  In the spirit of this result
we should also note an interesting introduction to the importance of the pure state
manifold $\mathbb{C}\mbox{P}^N$ for large $N$, especially with regard to
quantum entanglement, can be found in \cite{Brody}.
\subsection{Volume of $SU(N)/(U(1))^{N-1}$}
Recall that the action of a group in the adjoint representation produces
interesting orbits; the manifolds of which are called
\textit{generalized flag manifolds}, and appear very often
in geometric quantization, density matrices, entangled states, etc.
These manifolds can be represented by the coset $SU(N)/U(1)^{N-1}$;
obviously then the volume of such manifolds are 
quite important to our work.  By using equations \eqref{Uvol} and (\ref{eq:msflagdef}) we can
write down the general volume for such a coset, 
${SU(N)}/{U(1)}_{{SU(2)}}\times {U(1)}_{{SU(3)}}\times \cdots
\times {U(1)}_{{SU(N)}}$, as 
\begin{align}
V\biggl(\frac{{SU(N)}}{{U(1)}_{\lambda_3}\times {U(1)}_{\lambda_8} \times
  \ldots \times {U(1)}_{\lambda_{N^2-1}}}\biggr)
&= \frac{V_{{SU(N)}}}{V_{{U(1)}_{\lambda_3}}\times V_{{U(1)}_{\lambda_8}}
  \times \ldots V_{{U(1)}_{\lambda_{N^2-1}}}} \nonumber \\
&= \frac{2^{\frac{N-1}{2}}\pi^{\frac{(N-1)(N+2)}{2}}\sqrt{N}\prod^{N-1}_{k=1}
\biggl(\frac{1}{k!}\biggr)}{2\pi * \sqrt{3}\pi * \ldots * \pi \sqrt{\frac{2N}{N-1}}} \nonumber \\
& = \frac{2^{\frac{N-1}{2}}\pi^{\frac{(N-1)(N+2)}{2}}\sqrt{N}\prod^{N-1}_{k=1}
\biggl(\frac{1}{k!}\biggr)}{\prod^{N-1}_{l=1}\pi \sqrt{\frac{2(l+1)}{l}}} \nonumber \\
& = \frac{\pi^{\frac{N(N-1)}{2}}\sqrt{N}\prod^{N-1}_{k=1}
\biggl(\frac{1}{k!}\biggr)}{\prod^{N-1}_{l=1} \sqrt{\frac{(l+1)}{l}}}.
\end{align}
But we know 
\begin{equation}
\prod^{N-1}_{l=1} \sqrt{\frac{(l+1)}{l}} = \sqrt{\frac{2}{1}} * \sqrt{\frac{3}{2}} * \ldots 
* \sqrt{\frac{N}{N-1}} = \sqrt{N}.
\end{equation}
Thus we can see that
\begin{equation}
\label{flagcosetvolume}
V\biggl(\frac{{SU(N)}}{{U(1)}_{\lambda_3}\times {U(1)}_{\lambda_8} \times
  \ldots \times {U(1)}_{\lambda_{N^2-1}}}\biggr) = \pi^{\frac{N(N-1)}{2}}\prod^{N-1}_{k=1}
\biggl(\frac{1}{k!}\biggr),
\end{equation}
which is in agreement with \cite{Boya2}.
This volume can also be generated via 
integrating the ``truncated'' Haar measure,
using the appropriate ranges and normalization
conditions, given in equation (\ref{eq:dghmeasure}) (done in detail in
\cite{Tilma2}).

For completeness, and with equation \eqref{stratified} in mind, we
should note the following:
\begin{equation}
\pi^{\frac{N(N-1)}{2}} \equiv \prod_{k=1}^{N-1}\pi^k.
\end{equation}
Thus equation \eqref{flagcosetvolume} has the following, equivalent 
representation
\begin{eqnarray}
V\biggl(\frac{{SU(N)}}{{U(1)}_{\lambda_3}\times {U(1)}_{\lambda_8} \times
\cdots \times {U(1)}_{\lambda_{N^2-1}}}\biggr) 
&=& \pi^{\frac{N(N-1)}{2}}\prod^{N-1}_{k=1}\biggl(\frac{1}{k!}\biggr)
= \prod_{k=1}^{N-1}\frac{\pi^k}{k!} \nonumber \\ 
&=& \prod_{k=1}^{N-1}V_{\mathbb{C}\text{P}^k}.
\end{eqnarray}
So the volume of our flag manifold is nothing more than the product of
the volumes of the complex projective space $\mathbb{C}\mbox{P}^k$
where $k\leq N$ \cite{Boya2}.  Notice also that as $N$ increases the volume of the flag
manifold approaches, but \textit{never equals}, zero (see equation
(\ref{eq:cpnobs})).  It is an
asymptotic limit which converges to zero from the left on
$\mathbb{R}^1$.  Thus, since one \textit{usually} chooses a non-zero
probability distribution on the $N-1$ symplex defining $d\mu$ (see
equation (\ref{eq:dmu})), we can
conclude that, as in the pure state case, the mixed state volume
measure will never equal zero \textit{unless $V_{\text{Symplex}}$ does}!


\section{Other $SU(N)$ and $U(N$) Coset Volumes}

Beyond the full pure and mixed state manifolds there are numerous other
sub-manifolds that are of interest in physics; the volumes of which
have already been calculated (see, for example, 
\cite{Boya2,Marinov2,Zyczkowski2,Marinov,VK,Sinolecka} and
references within).  These sub-manifolds and their volumes 
give us both a way to confirm our
methodology, as well as offering a systematic,
rather than numeric, way of computing such quantities.
From this work we will then be able to calculate the
manifolds that contain the set of entangled and mixed states (either
pure or mixed) for specific quantum systems \cite{Tilma3}.  
It should be understood though that 
the following volume calculations are specific
to the $SU(N)$ and $U(N)$ Euler angle parameterization that we have
developed and 
its corresponding normalizations via the Cartan subalgebra being used.
The general question of volume normalization of a manifold, 
especially when one begins to talk
about coset manifolds with specific elements of the Cartan subalgebra being
removed will be the subject of a future paper.
\subsection{Volume of $SU(N)/SU(P) \times SU(Q)$}
To begin, we would like to be able
to write down the general volume of the ${SU(N)}/{SU(P)}\times {SU(Q)}$ coset
where $N+1 \geq P+Q $ and $P, Q \neq 1$.
To do this we can use equation \eqref{eulermarinov} to generate
\begin{align}
\frac{V_{SU(N)}}{V_{SU(P)} \times V_{SU(Q)}} &= \frac{2^{\frac{N-1}{2}}\pi^{\frac{(N-1)(N+2)}{2}}\sqrt{N}\prod^{N-1}_{k=1}
\biggl(\frac{1}{k!}\biggr)}{2^{\frac{P-1}{2}}\pi^{\frac{(P-1)(P+2)}{2}}\sqrt{P}\prod^{P-1}_{k=1}
\biggl(\frac{1}{k!}\biggr) \times 2^{\frac{Q-1}{2}}\pi^{\frac{(Q-1)(Q+2)}{2}}\sqrt{Q}\prod^{Q-1}_{k=1}
\biggl(\frac{1}{k!}\biggr)} \nonumber \\
&=2^{\frac{(N+1)-(P+Q)}{2}}\pi^{\frac{N(N+1)-P(P+1)-Q(Q+1)+2}{2}}\sqrt{\frac{N}{PQ}}\prod^{N-1}_{k=1}\biggl(\frac{1}{k!}\biggr)\prod^{P-1}_{k=1}{k!}
\prod^{Q-1}_{k=1}{k!}.
\label{eq:volsunsupsuq}
\end{align}
When $N+1=P+Q$ we have
\begin{align}
\frac{V_{SU(N)}}{V_{SU(P)} \times V_{SU(Q)}} &=
2^{\frac{(N+1)-(N+1)}{2}}\pi^{\frac{(P+Q-1)(P+Q)-P(P+1)-Q(Q+1)+2}{2}} \nonumber \\ 
&\times \sqrt{\frac{P+Q-1}{PQ}}\prod^{P+Q-2}_{k=1}\biggl(\frac{1}{k!}\biggr)\prod^{P-1}_{k=1}{k!}\prod^{Q-1}_{k=1}{k!} \nonumber \\
&= \pi^{(P-1)(Q-1)}\sqrt{\frac{P+Q-1}{PQ}}\prod^{P+Q-2}_{k=1}\biggl(\frac{1}{k!}\biggr)\prod^{P-1}_{k=1}{k!}\prod^{Q-1}_{k=1}{k!}
\end{align}
\subsubsection{Example Calculation: Volume of $SU(4)/SU(2) \times SU(2)$}
Defining $N=4$, and $P=Q=2$, we get from equation
(\ref{eq:volsunsupsuq}) the volume of the coset $SU(4)/SU(2) \times SU(2)$
\begin{align}
\frac{V_{SU(4)}}{V_{SU(2)} \times V_{SU(2)}} &=
2^{\frac{5-4}{2}}\pi^{\frac{4(5)-2(3)-2(3)+2}{2}}\sqrt{\frac{4}{(2)(2)}}\prod^{4-1}_{k=1}\biggl(\frac{1}{k!}\biggr)\prod^{2-1}_{k=1}{k!}\prod^{2-1}_{k=1}{k!}
\nonumber \\
&= \sqrt{2} * \pi^5 * \frac{1}{3!} * 1! * 1!
\nonumber \\
&= \frac{\pi^5}{6\sqrt{2}}.
\end{align}
Which is equivalent to the volume of $SU(4)$, $\sqrt{2}\pi^9/3$, divided
by the square of the volume of $SU(2)$, $2\pi^2$, as expected.  It should also be noted that this is the volume of the manifold that is 
comprised of all \textit{non-local} transformations which can be implemented on a two qubit system.
\subsection{Volume of ${SU(N)}/{U(P)}\times {U(1)}$}
Beyond the general volume of $\mathbb{C}\mbox{P}^N$, general flag manifold, and the previous $SU(N)$
coset, we would like to be able
to write down the general volume of the ${SU(N)}/{U(P)}\times {U(1)}$ coset
where $N-1 \geq P+1$  and ${P}\neq 1$.
To do this we can use equations \eqref{eulermarinov} and
(\ref{eq:UNvol}) as follows
\begin{equation}
\frac{V_{{SU(N)}}}{V_{{U(P)}}\times V_{{U(1)}}} = \frac{2^{\frac{N-1}{2}}\pi^{\frac{(N-1)(N+2)}{2}}\sqrt{N}\prod^{N-1}_{k=1}
\biggl(\frac{1}{k!}\biggr)}{2^{\frac{P}{2}}\pi^{\frac{P(P+1)}{2}}\sqrt{P+1}\prod^{P-1}_{k=1}
\biggl(\frac{1}{k!}\biggr)\times V_{{U(1)}}}.
\end{equation}
The problem we now face is how to define $U(1)$.  If we use equation
\eqref{Uvol}, here now defined for $SU(N)$, we would generate
\begin{align}
\frac{V_{{SU(N)}}}{V_{{U(P)}}\times V_{{U(1)}_{{SU(N)}}}} &= \frac{2^{\frac{N-1}{2}}\pi^{\frac{(N-1)(N+2)}{2}}\sqrt{N}\prod^{N-1}_{k=1}\biggl(\frac{1}{k!}\biggr)}{2^{\frac{P}{2}}\pi^{\frac{P(P+1)}{2}}\sqrt{P+1}\prod^{P-1}_{k=1}\biggl(\frac{1}{k!}\biggr)* \pi\sqrt{\frac{2N}{N-1}}} \nonumber \\
&= \frac{2^{\frac{N-1}{2}}\pi^{\frac{(N-1)(N+2)}{2}}\sqrt{N-1}\prod^{N-1}_{k=1}\biggl(\frac{1}{k!}\biggr)}
{2^{\frac{P+1}{2}}\pi^{\frac{P^2+P+2}{2}}\sqrt{P+1}\prod^{P-1}_{k=1}
\biggl(\frac{1}{k!}\biggr)} \nonumber \\
&=2^{\frac{(N-1)-(P+1)}{2}}\pi^{\frac{(N^2+N-2)-(P^2+P+2)}{2}}\sqrt{\frac{N-1}{P+1}}\prod^{N-1}_{k=1}\biggl(\frac{1}{k!}\biggr){\prod^{P-1}_{k=1}{k!}}.
\end{align}
If we demand that $N-1 = P+1$, we can simplify the product terms
\begin{align}
\prod^{N-1}_{k=1}\biggl(\frac{1}{k!}\biggr)\prod^{P-1}_{k=1}{k!}
&= \prod^{P+1}_{k=1}\biggl(\frac{1}{k!}\biggr)\prod^{P-1}_{k=1}{k!}
\nonumber \\
&= \frac{1!*2!*\cdots*(P-2)!*(P-1)!}{1!*2!*\cdots*(P-2)!*(P-1)!*P*(P+1)!}
\nonumber \\
&= \frac{1}{P!(P+1)!}
\end{align}
as well as the powers and other factors.  Therefore, for this case we have
\begin{align}
\frac{V_{{SU(N)}}}{V_{{U(P)}}\times V_{{U(1)}_{{SU(N)}}}} &=
\frac{2^{\frac{(N-1)-(P+1)}{2}}\pi^{\frac{(N^2+N-2)-(P^2+P+2)}{2}}}{P!(P+1)!}\sqrt{\frac{N-1}{P+1}}
\nonumber \\
&=\frac{\pi^{2N-3}}{(N-2)!(N-1)!} \nonumber \\
&=\frac{\pi^{2P+1}}{P!(P+1)!}.
\label{eq:firstcosetvol}
\end{align}
Depending on which parameter is used. 

Now, if in using equation \eqref{Uvol}, we now define $U(1)$ for $SU(M)$,
$M<N$, we would generate
\begin{align}
\frac{V_{{SU(N)}}}{V_{{U(P)}}\times V_{{U(1)}_{SU(M)}}} &=
\frac{2^{\frac{N-1}{2}}\pi^{\frac{(N-1)(N+2)}{2}}\sqrt{N}\prod^{N-1}_{k=1}\biggl(\frac{1}{k!}\biggr)}{2^{\frac{P}{2}}\pi^{\frac{P(P+1)}{2}}\sqrt{P+1}\prod^{P-1}_{k=1}\biggl(\frac{1}{k!}\biggr)*\pi\sqrt{\frac{2M}{M-1}}} \nonumber \\
&=2^{\frac{(N-1)-(P+1))}{2}}\pi^{\frac{(N^2+N-2)-(P^2+P+2)}{2}}\sqrt{\frac{N(M-1)}{M(P+1)}}\prod^{N-1}_{k=1}\biggl(\frac{1}{k!}\biggr){\prod^{P-1}_{k=1}{k!}},
\end{align}
and if we demand $N-1 = P+1$, we can simplify, yielding
\begin{align}
&=
\frac{2^{\frac{(N-1)-(P+1))}{2}}\pi^{\frac{(N^2+N-2)-(P^2+P+2)}{2}}}{P!(P+1)!}\sqrt{\frac{N(M-1)}{M(P+1)}}\nonumber \\
&=\frac{\pi^{2N-3}}{(N-2)!(N-1)!}\sqrt{\frac{N(M-1)}{M(N-1)}},
\label{eq:secondcosetvol}
\end{align}
which reduces to equation (\ref{eq:firstcosetvol}) when $M=N$.  Therefore,
depending on which $U(1)$ we use, we will generate a different volume; the
ratio between any two being equal to
\begin{equation}
\frac{V_{{SU(N)}}/V_{{U(P)}}\times V_{{U(1)}_{SU(X)}}}
{V_{{SU(N)}}/V_{{U(P)}}\times V_{{U(1)}_{SU(Y)}}} = \sqrt{\frac{Y(X-1)}{X(Y-1)}}.
\end{equation}
\subsubsection{Example Calculation: Volumes of $SU(4)/U(2) \times 
U(1)_{SU(i)}$ for $i=2,3,4$}
Defining $N=4$ and $P=2$ (thus satisfying $N-1=P+1$ we get from equation
(\ref{eq:secondcosetvol}) the volume of the coset $SU(4)/U(2) \times
U(1)_{SU(i)}$ when $i=2$
\begin{equation}
\frac{V_{{SU(4)}}}{V_{{U(2)}}\times V_{{U(1)}_{{SU(2)}}}} =
\frac{\pi^{2*4-3}}{(4-2)!(4-1)!}\sqrt{\frac{4(2-1)}{2(4-1)}}
= \frac{\pi^5}{12}\sqrt{\frac{2}{3}}
= \frac{\pi^5}{6\sqrt{6}}\,,
\end{equation}
when $i=3$
\begin{equation}
\frac{V_{{SU(4)}}}{V_{{U(2)}}\times V_{{U(1)}_{{SU(3)}}}} =
\frac{\pi^{2*4-3}}{(4-2)!(4-1)!}\sqrt{\frac{4(3-1)}{3(4-1)}}
= \frac{\pi^5}{12}\sqrt{\frac{8}{9}}
= \frac{\pi^5}{9\sqrt{2}}\,,
\end{equation}
and when $i=4$ the  volume of the coset $SU(4)/U(2) \times
U(1)_{SU(4)}$, using equation (\ref{eq:firstcosetvol}) now, is
\begin{equation}
\frac{V_{{SU(4)}}}{V_{{U(2)}}\times V_{{U(1)}_{{SU(4)}}}} =
\frac{\pi^{2*4-3}}{(4-2)!(4-1)!}
= \frac{\pi^5}{12}\,.
\end{equation}
\subsection{Volume of ${SU(N)}/{U(P)}\times {U(Q)}$}
Now we would like to be able
to write down the general volume of the ${SU(N)}/{U(P)}\times {U(Q)}$ coset
for $N-1 \geq P+Q $ and $P, Q \neq 1$.
To do this we can use equations \eqref{eulermarinov} and
(\ref{eq:UNvol}) as follows
\begin{equation}
\frac{V_{{SU(N)}}}{V_{{U(P)}}\times V_{{U(Q)}}} = \frac{2^{\frac{N-1}{2}}\pi^{\frac{(N-1)(N+2)}{2}}\sqrt{N}\prod^{N-1}_{k=1}
\biggl(\frac{1}{k!}\biggr)}{2^{\frac{P}{2}}\pi^{\frac{P(P+1)}{2}}\sqrt{P+1}\prod^{P-1}_{k=1}
\biggl(\frac{1}{k!}\biggr)\times2^{\frac{Q}{2}}\pi^{\frac{Q(Q+1)}{2}}\sqrt{Q+1}\prod^{Q-1}_{k=1}
\biggl(\frac{1}{k!}\biggr)}.
\end{equation}
Simplification yields
\begin{eqnarray}
\frac{V_{{SU(N)}}}{V_{{U(P)}}\times V_{{U(Q)}}} &=&
\frac{2^{\frac{N-1}{2}}\pi^{\frac{(N-1)(N+2)}{2}}\sqrt{N}\prod^{N-1}_{k=1}\biggl(\frac{1}{k!}\biggr)}{2^{\frac{P+Q}{2}}\pi^{\frac{P(P+1)+Q(Q+1)}{2}}\sqrt{(P+1)(Q+1)}\prod^{P-1}_{k=1}\biggl(\frac{1}{k!}\biggr)\prod^{Q-1}_{k=1}\biggl(\frac{1}{k!}\biggr)}
\nonumber \\
&=& 2^{\frac{(N-1)-(P+Q)}{2}}\pi^{\frac{(N-1)(N+2)-P(P+1)-Q(Q+1)}{2}}\sqrt{\frac{N}{(P+1)(Q+1)}} \nonumber \\
&&\times \prod^{N-1}_{k=1}\biggl(\frac{1}{k!}\biggr)\prod^{P-1}_{k=1}k!\prod^{Q-1}_{k=1}k!.
\end{eqnarray}
For the special case when $N-1 = P+Q$ we can go further and 
eliminate the $N$ dependence in the above
volume, thus yielding (in one possible representation)
\begin{eqnarray}
\frac{V_{{SU(N)}}}{V_{{U(P)}}\times V_{{U(Q)}}} &=&
\pi^{(P+Q+PQ)}\sqrt{\frac{P+Q+1}{(P+1)(Q+1)}}\prod^{P+Q}_{k=1}\biggl(\frac{1}{k!}\biggr)\prod^{P-1}_{k=1}k!\prod^{Q-1}_{k=1}k!
\nonumber \\
&=&
\pi^{(P+Q+PQ)}\sqrt{\frac{P+Q+1}{(P+1)(Q+1)}}\prod^{P+Q}_{k=P}\biggl(\frac{1}{k!}\biggr)\prod^{Q-1}_{k=1}k!.
\label{eq:sunupuqcoset}
\end{eqnarray}
\subsubsection{Example Calculation: Volume of $SU(9)/U(4) \times U(4)$}
Defining $N=9$, and $P=Q=4$, thus satisfying $N-1=P+Q$, we get from
equation (\ref{eq:sunupuqcoset}) the volume of the coset $SU(9)/U(4)
\times U(4)$ to be equal to
\begin{align}
\frac{V_{{SU(9)}}}{V_{{U(4)}}\times V_{{U(4)}}} &=
\pi^{(4+4+4*4)}\sqrt{\frac{4+4+1}{(4+1)(4+1)}}\prod^{4+4}_{k=4}\biggl(\frac{1}{k!}\biggr)\prod^{4-1}_{k=1}k! 
\nonumber \\
&=\frac{\pi^{24}}{58525286400000},
\end{align}
which is what one would get if they used equations
\eqref{eulermarinov} and (\ref{eq:UNvol}) separately.
\subsection{Volume of ${SU(N)}/\prod_{i=1}^x {U}(P_i)\times \prod_{j=1}^y {U(1)}_{{SU}(Z_j)}$}
We are now ready to write down the volume for the most general of
cosets that we are interested in, ${SU(N)}/\prod_{i=1}^x {U}(P_i)\times \prod_{j=1}^y {U(1)}_{{SU}(Z_j)}$,
where
\begin{equation}
\label{mostgencond}
\sum_{i=1}^x P_i + \sum_{j=1}^y 1 = \sum_{i=1}^x P_i + y \leq N-1, \quad P_i\neq 1,
\end{equation}
and 
\begin{equation}
{U(1)}_{{SU}(Z_j)} \in \{{U(1)}_{{SU(2)}}, {U(1)}_{{SU(3)}},\ldots, {U(1)}_{{SU(N)}}\},
\end{equation}
where there is no necessary order in the sequential choice of
${U(1)}_{{SU}(Z_j)}$.

To begin we note the following using equation (\ref{eq:UNvol})
\begin{align}
V(\prod_{i=1}^x {U}(P_i)) &= \prod_{i=1}^x V_{{U}(P_i)} \nonumber \\
&= \prod_{i=1}^x \biggr(\;2^{\frac{P_i}{2}}\pi^{\frac{P_i(P_i+1)}{2}}\sqrt{P_i+1}\prod^{P_i-1}_{k=1}
\biggl(\frac{1}{k!}\biggr)\biggl) \nonumber \\
&=2^{\frac{\sum_{i=1}^x P_i}{2}}\pi^{\frac{\sum_{i=1}^x P_i(P_i+1)}{2}}\prod_{i=1}^x \biggr(\sqrt{P_i+1}
\prod^{P_i-1}_{k=1}\biggl(\frac{1}{k!}\biggr)\biggl).
\label{eq:denom1}
\end{align}
We also can simplify the second volume of the three we need via
generalizing equation \eqref{Uvol}
\begin{align}
V(\prod_{j=1}^y {U(1)}_{{SU}(Z_j)}) &= \prod_{j=1}^y V_{{U(1)}_{{SU}(Z_j)}}
= \prod_{j=1}^y \pi\sqrt{\frac{2Z_j}{Z_j-1}} \nonumber \\
&=\; \pi^y 2^{\frac{y}{2}}\prod_{j=1}^y \sqrt{\frac{Z_j}{Z_j-1}}.
\label{eq:denom2}
\end{align}
We are now in a position to write down the volume for
${SU(N)}/\prod_{i=1}^x {U}(P_i)\times \prod_{j=1}^y {U(1)}_{{SU}(Z_j)}$.
Using equations \eqref{eulermarinov}, \eqref{mostgencond}, (\ref{eq:denom1}), and
(\ref{eq:denom2}) we have
\begin{eqnarray}
\label{bigeqn}
V\biggr(\frac{{SU(N)}}{\prod_{i=1}^x{U}(P_i) \times \prod_{j=1}^y
{U(1)}_{{SU}(Z_j)}}\biggl)
&=& 2^{\frac{N-(1+y+\sum_{i=1}^x P_i)}{2}} \pi^{\frac{(N-1)(N+2)-(2y+\sum_{i=1}^x P_i(P_i+1))}{2}} \nonumber \\
&&\times \sqrt{N} \frac{\prod^{N-1}_{k=1}\biggl(\frac{1}{k!}\biggr)\prod_{j=1}^y \sqrt{\frac{Z_j-1}{Z_j}}}
{\prod_{i=1}^x\biggr(\sqrt{P_i+1}\prod^{P_i-1}_{k=1}\biggl(\frac{1}{k!}\biggr)\biggl)}.
\end{eqnarray}
For the special case when the ``$\leq$'' in equation \eqref{mostgencond}
is replaced by ``$=$'' we have
\begin{eqnarray}
V\biggr(\frac{{SU(N)}}{\prod_{i=1}^x{U}(P_i) \times \prod_{j=1}^y
{U(1)}_{{SU}(Z_j)}}\biggl) &=& \pi^{\frac{(N-1)(N+2)-(2y+\sum_{i=1}^x P_i(P_i+1))}{2}}\\
&&\times \sqrt{N}\; \frac{\prod^{N-1}_{k=1}\biggl(\frac{1}{k!}\biggr)\prod_{j=1}^y \sqrt{\frac{Z_j-1}{Z_j}}}
{\prod_{i=1}^x\biggr(\sqrt{P_i+1}\prod^{P_i-1}_{k=1}\biggl(\frac{1}{k!}\biggr)\biggl)}. \nonumber
\end{eqnarray}
One could continue simplifying equation \eqref{bigeqn} but it would be
only worthwhile if additional knowledge concerning $Z_j$ and $P_i$ 
was available.
\subsection{Grassmann Volume}
The general Grassmann manifolds, of which $\mathbb{C}\mbox{P}^N$
is a special case (see equation \eqref{grasscpn}), have the following
definition for $N\geq M$
\begin{equation}
G(N,M) 
= \frac{{U(N)}}{{U(M)}\times {U(N-M)}}.
\end{equation}
Using equation (\ref{eq:UNvol}) we can write down the general
expression for the volume of \textit{almost} any Grassmann manifold:
\begin{align}
&V_{G(N,M)} = \frac{V_{U(N)}}{V_{U(M)} \times V_{U(N-M)}} \nonumber \\
&=\frac{2^{\frac{N}{2}}\pi^{\frac{N(N+1)}{2}}\sqrt{N+1}\prod^{N-1}_{k=1}
\biggl(\frac{1}{k!}\biggr)}{2^{\frac{M}{2}}\pi^{\frac{M(M+1)}{2}}\sqrt{M+1}\prod^{M-1}_{k=1}\biggl(\frac{1}{k!}\biggr)
\times
\;2^{\frac{N-M}{2}}\pi^{\frac{(N-M)(N-M+1)}{2}}\sqrt{N-M+1}\prod^{N-M-1}_{k=1}\biggl(\frac{1}{k!}\biggr)}
\nonumber \\
&=\pi^{M(N-M)}\sqrt{\frac{N+1}{(M+1)(N-M+1)}}\prod^{N-1}_{k=1}
\biggl(\frac{1}{k!}\biggr)\prod^{M-1}_{k=1}k!\prod^{N-M-1}_{k=1}k!
\nonumber \\
&=\pi^{M(N-M)}\sqrt{\frac{N+1}{(M+1)(N-M+1)}}\prod^{N-1}_{k=M}
\biggl(\frac{1}{k!}\biggr)\prod^{N-M-1}_{k=1}k!.
\label{eq:grassvol}
\end{align}
The reason for the ``almost'' above is that for $M=1$ we \text{do not} 
regain the volume for $\mathbb{C}P^{N-1}$ that we originally
calculated in equation (\ref{eq:CPNvol}):
\begin{align}
V_{\mathbb{C}\text{P}^{N-1}} \equiv V_{G(N,1)} &= \pi^{N-1}\sqrt{\frac{N+1}{(1+1)(N-1+1)}}\prod^{N-1}_{k=1}
\biggl(\frac{1}{k!}\biggr)\prod^{N-1-1}_{k=1}k! \nonumber \\
&= \pi^{N-1}\sqrt{\frac{N+1}{2N}}\prod^{N-1}_{k=1}
\biggl(\frac{1}{k!}\biggr)\prod^{N-2}_{k=1}k! \nonumber \\
&= \frac{\pi^{N-1}}{(N-1)!}\sqrt{\frac{N+1}{2N}} \nonumber \\
&\neq \frac{\pi^{N-1}}{(N-1)!}.
\label{eq:cpnwrongvol}
\end{align}
We are ``off'' by a factor of $\sqrt{(N+1)/2N}$ which occurs because
of the following reason: equation (\ref{eq:UNvol}), for $N=1$, yields
$2\pi$ which is correct \textit{if} one is looking for the volume of
the $SU(2)$ variant of $U(1)$ (see equation \eqref{Uvol} for $N=1$),
but that is \textit{not} the case for the $U(1)$ components of 
greater $SU(N)$ groups (again see equation \eqref{Uvol} for $N\geq
2$), which is the case here.  In equation (\ref{eq:cpnwrongvol}) we get the
factor of $2\pi$ from the $U(1)$ component, but without the additional
contraction term due to the $\lambda_{N^2-1}$ Cartan subalgebra
component of $U(N)$ from which the $U(1)$ term is defined!

The flaw in equation (\ref{eq:cpnwrongvol}) can also be seen from equation 
\eqref{grasscpn}:
\begin{equation}
\mathbb{C}\mbox{P}^{N-1}\equiv G(N,1) 
= \frac{{U(N)}}{{U(1)}\times {U(N-1)}} = \frac{SU(N) \times U(1)}{U(1)
  \times U(N-1)} = \frac{{SU(N)}}{{U(N-1)}}.
\end{equation}
The $U(1)$ term in the numerator is the same as the $U(1)$ term in the 
denominator and as such, in any representation, cancels out, thus
leaving the standard coset relationship for $\mathbb{C}\mbox{P}^{N-1}$
which, from equation (\ref{eq:CPNvol}), does yield the correct volume
for $\mathbb{C}\mbox{P}^{N-1}$.  Therefore, 
if we use equation \eqref{Uvol} for the $U(1)$ term in equation
(\ref{eq:grassvol}) (combined with equation (\ref{eq:UNvol}) for the 
other two terms) when $M=1$, we will generate the correct volume
for $\mathbb{C}\mbox{P}^{N-1}$
\begin{align}
V_{G(N,1)} &= \frac{V_{U(N)}}{V_{U(1)} \times V_{U(N-1)}} = \frac{2^{\frac{N}{2}}\pi^{\frac{N(N+1)}{2}}\sqrt{N+1}\prod^{N-1}_{k=1}
\biggl(\frac{1}{k!}\biggr)}{ \pi\sqrt{\frac{2(N+1)}{N}} \times
\;2^{\frac{N-1}{2}}\pi^{\frac{(N-1)(N)}{2}}\sqrt{N}\prod^{N-2}_{k=1}\biggl(\frac{1}{k!}\biggr)}\nonumber
\\
&= \frac{\pi^{N-1}}{(N-1)!}.
\end{align}
Therefore, in general, if we demand that $M\neq 1$ then equation (\ref{eq:grassvol}) will
correctly produce the Grassmann manifold volumes (see \cite{Boya2} and
references within).


\section{Conclusion}

Using the volume equations given herein, we are now in a position to explicitly 
write down the measures and volumes for the whole range of manifolds which occur
in discussions concerning separability and entanglement of multi-particle systems.
Therefore, this work allows us to explicitly write down the volumes of the 
manifolds of the local orbits of a given state $\ket{\psi}$ with respect to 
some transformation $U \in SU(N)$ (or more generally $U(N)$), in a manner that 
we hypothesize also elucidates the topology of the manifolds as well \cite{Boya2,Sinolecka}.
Applications beyond quantum information theory are also possible \cite{Boya2}.


\section*{Acknowledgments}

We would like to thank Dr. Luis Boya for his informative discussions and calculations
with regards to the development of this paper.

\end{document}